\documentstyle[aps,preprint]{revtex}
\input{epsf}
\newcommand{\slL}{\raise.15ex\hbox{$/$}\kern-.53em\hbox{$L$}}
\newcommand{\slP}{\raise.15ex\hbox{$/$}\kern-.53em\hbox{$P$}}
\newcommand{\slR}{\raise.15ex\hbox{$/$}\kern-.53em\hbox{$R$}}
\newcommand{\slQ}{\raise.15ex\hbox{$/$}\kern-.53em\hbox{$Q$}}

\font\cmr=cmr7
\font\cmrX=cmr10

\font\tenimbf=cmmib10 at 12pt
\font\sevenimbf=cmmib10 at 7pt
\font\fiveimbf=cmmib10 at 5pt

\newfam\imbf
\textfont\imbf=\tenimbf
\scriptfont\imbf=\sevenimbf
\scriptscriptfont\imbf=\fiveimbf
\def\imb{\fam\imbf\tenimbf}

\font\tenmsa=msam10
\font\sevenmsa=msam7
\font\fivemsa=msam5
\font\tenmsb=msbm10
\font\sevenmsb=msbm7
\font\fivemsb=msbm5
\newfam\msafam
\newfam\msbfam
\textfont\msafam=\tenmsa  \scriptfont\msafam=\sevenmsa
  \scriptscriptfont\msafam=\fivemsa
\textfont\msbfam=\tenmsb  \scriptfont\msbfam=\sevenmsb
  \scriptscriptfont\msbfam=\fivemsb
\catcode`\@=11

\def\Bbb{\ifmmode\let\next\Bbb@\else
 \def\next{\errmessage{Use \string\Bbb\space only in math mode}}\fi\next}
\def\Bbb@#1{{\Bbb@@{#1}}}
\def\Bbb@@#1{\fam\msbfam#1}
\catcode`\@=12

\begin{document}
\begin{titlepage}
\title{\bf{Enhanced photon production rate\\
on the light--cone}}
\author{
P.~Aurenche$^{*,(1),(2)}$, F.~Gelis$^{(2)}$, 
R.~Kobes$^{(3)}$, E.~Petitgirard$^{(3)}$}
\address{\begin{itemize}
\begin{enumerate}
\item CFIF, Instituto Superior T\'ecnico, Edificio Ci\^enca (f\'\i sica),
P-1096 LISBOA, Codex, Portugal
\item Laboratoire de Physique Th\'eorique ENSLAPP,
B.~P.~110,  F--74941 Annecy-le-Vieux Cedex, France
\item Physics Department and Winnipeg Institute
for Theoretical Physics,
University of Winnipeg,
Winnipeg, Manitoba R3B 2E9, Canada
\end{enumerate}
\end{itemize}\vspace*{10pt}
\hfill{\it In memory of Tanguy Altherr}}
\maketitle
\begin{abstract}
Recent studies of the high temperature soft photon production
rate on the light--cone using Braaten--Pisarski resummation techniques
have found collinear divergences present. We show that there exist
a class of terms outside the Braaten--Pisarski framework which,
although also divergent, dominate over these previously considered
terms. The divergences in these new terms may be alleviated by
application of a recently developed resummation scheme
for processes sensitive to the light--cone.
\end{abstract}
\begin{flushright}
ENSLAPP--A--586/96\\
FISIST/4--96/CFIF\\
WIN--96--5\\
\end{flushright}
\vfill
\hbox to 3cm{\hrulefill}
\vglue 3mm
\hbox to 10cm{\cmrX $^*$ On leave of absence from ENSLAPP, B.P. 110, 
F--74941 Annecy-le-Vieux Cedex, France\hfil}
\thispagestyle{empty}
\end{titlepage}
\section{Introduction}
The development by Braaten and Pisarski
of the effective expansion of hot gauge theories
\cite{rp,brat}, given in terms
of hard thermal loops \cite{klim,wel,wong,tay},
has resolved some long--standing
paradoxes in the field \cite{rolf,thoma,rev}.
However, it is also realized that these techniques are
useful down to scales of the external momenta of the
order of $gT$, the ``soft'' scale; infrared problems
still remain if one goes below this scale, such as in
calculations involving the fast fermion damping rate
\cite{smilga,cliff,reba,alt,piz}, the calculation
of corrections to the Debye mass \cite{joe,debye,nieto},
and calculations of the $QCD$ pressure
\cite{pressure,eff1,eff2}. As well, problems also
arise for processes
sensitive to the behaviour of the theory near the light
cone, such as in the soft photon production
rate \cite{schiff,pat,peter}, photon bremsstrahlung from
a $QED$ or $QCD$ plasma \cite{weldon,cleymans}, 
including the effect of
Landau-Pomeranchuk suppression \cite{lanpom},
or scalar $QED$ and $QCD$ dispersion
relations \cite{kraemmer,flech,schulz}.
All of these problems
warrant extensions of the hard thermal loop resummation
techniques, although it is not obvious whether or
not such extensions will be perturbative in nature.
\par
In this note we consider the production of a real photon with 
momentum of $\cal O$($g T$). Concerning the calculation of 
this rate the following paradox appears: a straightforward
application of the hard thermal loop (HTL)
effective expansion leads to a rate
of $\cal O$($e^2 g^3 T^2 $) (neglecting logarithmic divergences)
and the production process is dominated by diagrams involving
soft fermions; on the other hand, the bremsstrahlung emission
of photons by hard (momentum of $\cal O$($T$)) fermions
has been estimated using semi-classical methods \cite{weldon,cleymans}
and it was found to be of $\cal O$($e^2 g T^2 $)
(ignoring the Landau-Pomeranchuck effect).
In the framework of the hard thermal loop expansion 
such  bremsstrahlung diagrams, involving  a hard fermion loop,
should be suppressed rather than enhanced.

We re-examine below the problem of the hard fermion loop 
contribution to the production of soft real photons in the
framework of thermal field theory, going beyond the hard loop 
expansion. We find that the sensitive behaviour of these terms
to the light cone enhances their order by a factor of
$1/g^2$ relative to the soft terms. As with the soft
loop contributions, these enhanced terms
also exhibit a collinear divergence, which however is
alleviated by including thermal mass effects on the hard
fermion propagators as required for a consistent calculation
and also in agreement with a recently developed
extension to the hard thermal loop resummation scheme
for processes near the light cone \cite{flech}.
Although a rigorous proof that this extension
is complete is still forthcoming,
it is clear that these terms dominate those of
the hard thermal loop effective expansion, and as such
a new effective expansion in cases such as this should
be investigated.

\section{Production rate}
\begin{figure}[ht]
\begin{center}
\leavevmode
\leftskip -1cm
\epsfbox{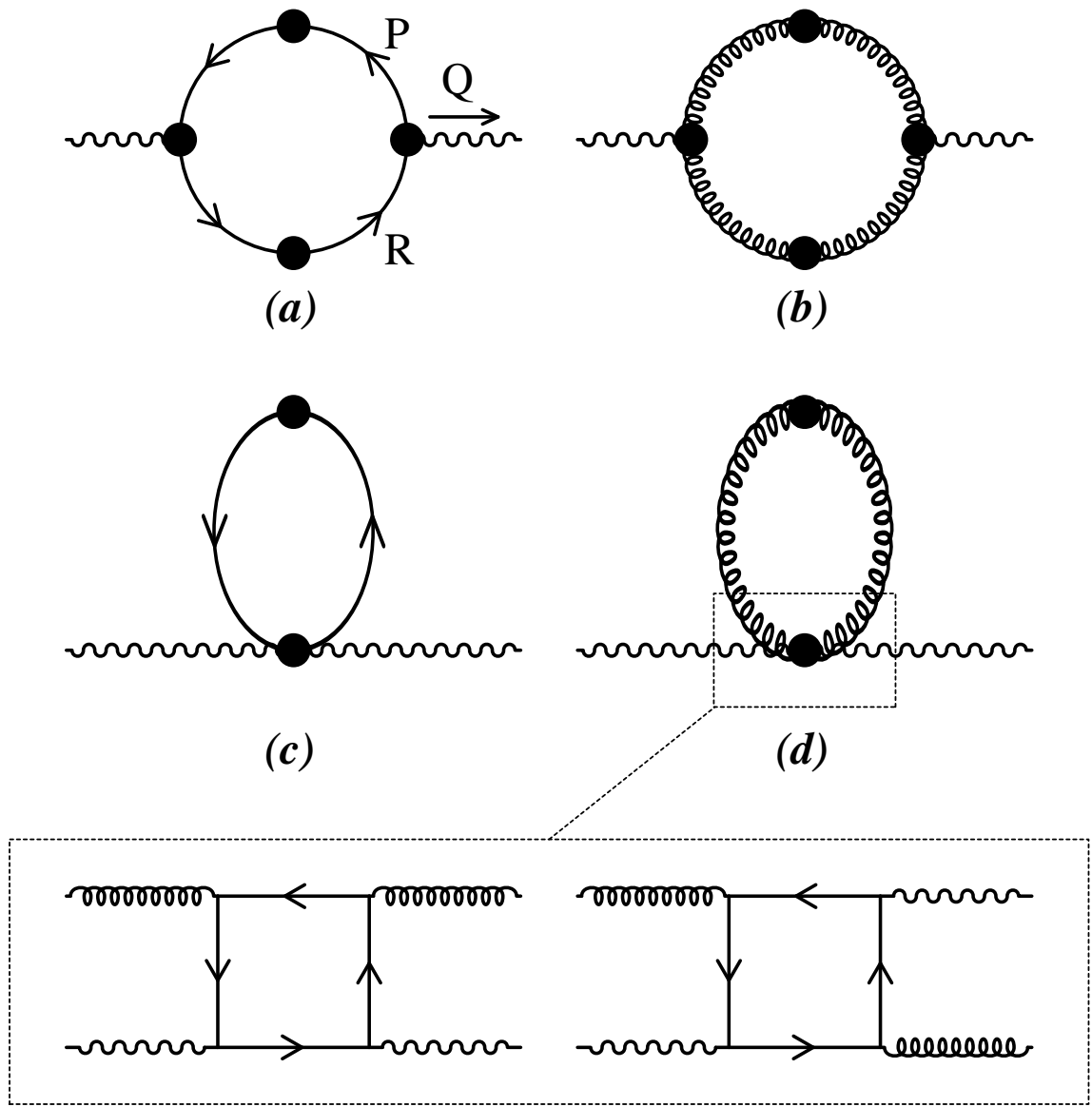}
\leftskip 0cm
\end{center}
\caption{Contributions to the soft photon production
rate with soft internal lines.}
\label{figsoft}\end{figure}

In order to calculate the photon production rate, we must evaluate the
imaginary part of the
trace of the (retarded) polarization tensor:
\begin{equation}
  \label{rate}
  q_0 \frac{d \sigma}{d^3 q} = - \frac{1 }{ (2 \pi)^3} n_{_{B}} (q_0)
  \ \hbox{\rm Im}\, \Pi^\mu\,_\mu (Q)
  \sim {1 \over g} \ \hbox{\rm Im}\, \Pi^\mu\,_\mu (Q)
\end{equation}
where the approximate equality holds for a photon of energy 
$q_0 \sim g T$. According to the Braaten-Pisarski theory, 
the four diagrams displayed in Fig.~\ref{figsoft} could,
a priori, contribute to soft photon production at leading-order.
However, the diagram 
of Fig.~\ref{figsoft}.b is zero thanks to an extension of
the Furry's theorem to the effective vertex with one photon and
two gluons. Moreover, in the HTL approximation, the contribution 
of the diagrams of
Fig.~\ref{figsoft}.c and Fig.~\ref{figsoft}.d are known to vanish:
indeed, the trace of the 4-point function with two photons and two
fermions vanishes while the 4-gauge-boson effective
vertex of Fig.~\ref{figsoft}.c has no HTL contribution. There remains
only the diagram of Fig.~\ref{figsoft}.a which is expected to contribute
a term of ${\cal O} \ (e^2 g^2 T^2)$ to $\hbox{\rm Im}\, \Pi^\mu\,_\mu (Q)$
since the loop momentum is soft and ``effective propagators and
vertices are of the same order as their bare counterparts" 
\cite{rp,brat}.
The contribution from this soft internal quark loop
using the effective propagators and vertices 
has been calculated in Refs.\cite{schiff,pat}, with a result that,
to leading order, exhibits a collinear divergence:
\begin{eqnarray}
 \hbox{\rm Im}\, \Pi^\mu\,_\mu(Q) &=& -4\ N_c e^2\
  \frac{m^2_{\hbox{\cmr th}}}{ q^2} \int{{d\hat{L}}\over
{(2\pi)^{1-2\varepsilon}}}{{q}\over{Q\cdot\hat{L}+i\epsilon}}
\int \frac{d^n P}{(2\pi)^{n-1}}\ 2\pi\delta(P\cdot\hat Q)
\nonumber\\
& &\;\; \times(\frac{1}{2}-n_{_{F}}(p_0))
 \sum_{s=\pm 1} \left[ (1-{sp_0 \over p}) \beta_s (P) +
  (1- {sr_0 \over r}) \beta_s (R)\right]\nonumber\\
  & & \;\;\sim O(e^2 g^3T^2)\frac{1}{\varepsilon},
\label{soft}
\end{eqnarray}
where $m^2_{\hbox{\cmr th}} = C_F g^2T^2 / 8$,
$P=(p_0,\vec p)$, $p=|\vec p|$, $\varepsilon$ is the
regulating parameter of dimensional regularization
($n=4+2\varepsilon\,, \varepsilon > 0$),
 and the effective quark propagator has been split as
\begin{equation}
^*S_{_{R}}(P) \equiv i \sum_{s=\pm 1}
\frac{\hat{\slP}_s}{D^s_{_{R}} (p_0 + i \epsilon, \vec p)},
\end{equation}
where $\hat{P}_s$ is the light-like vector
$\hat{P}_s \equiv (1,s\hat{p}=s{\vec p}/p)$ and
$1/ D^s_{_{R}} \equiv \alpha_s(P) - i \pi \beta_s (P)$.
\par
\begin{figure}[ht]
\begin{center}
\leavevmode
\epsfxsize=5 in
\vbox{\epsfbox{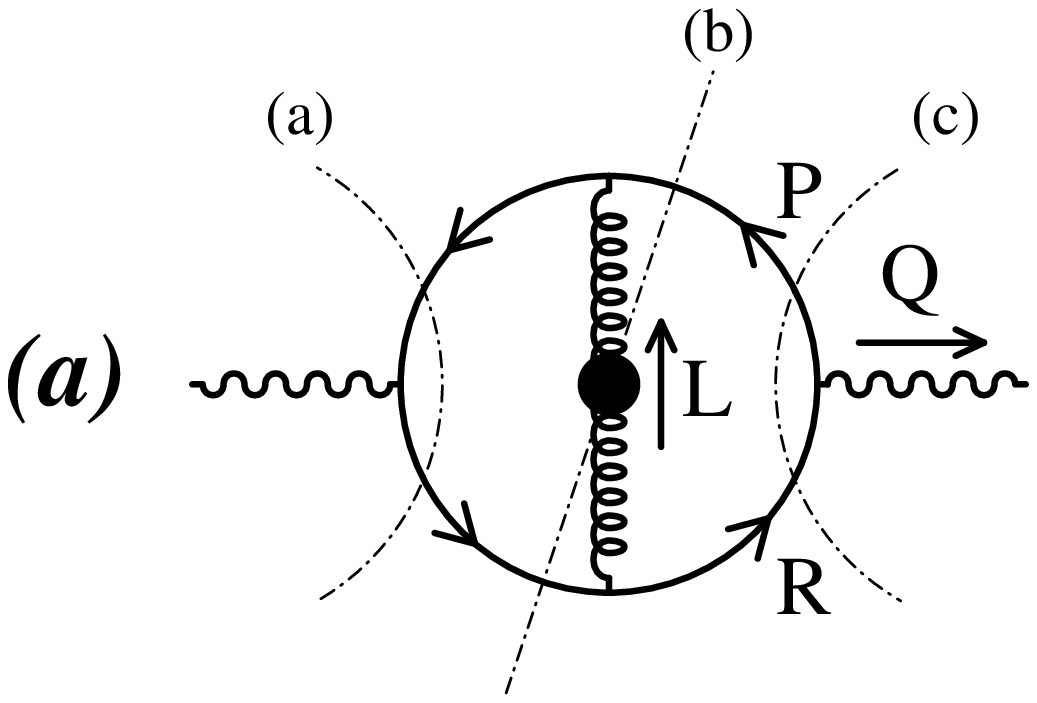}\epsfbox{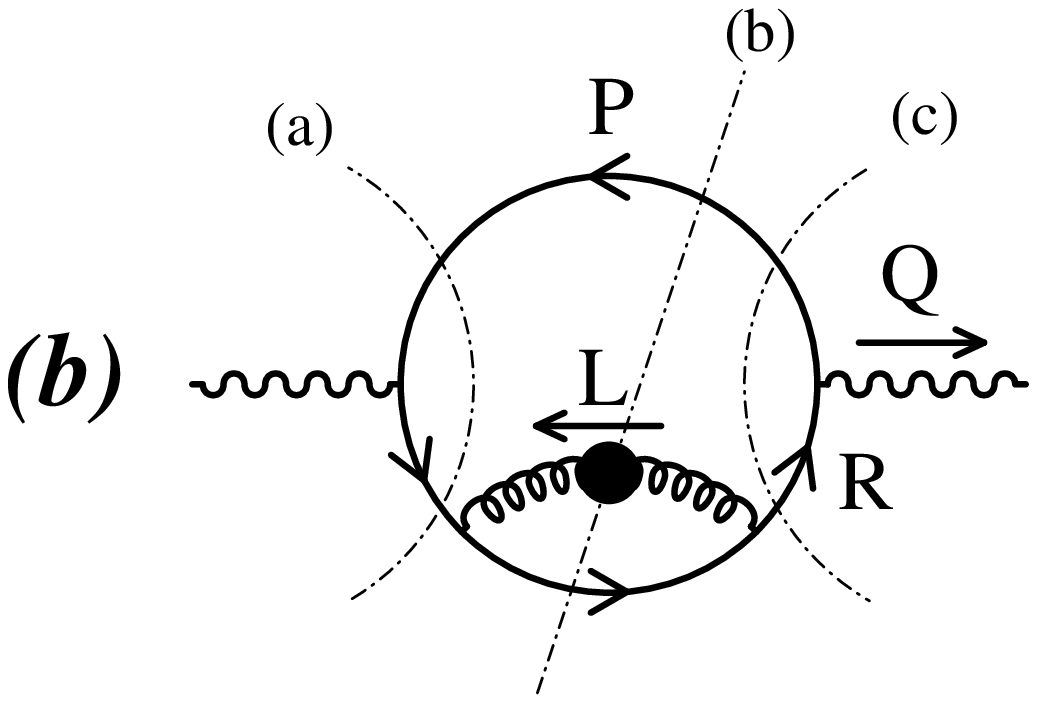}}
\end{center}
\caption{Contributions to the soft photon production
rate with hard internal fermion lines -- (a): vertex insertion;
(b): self--energy insertion.}
\label{fighard}\end{figure}

One should note that the above result is suppressed by a factor
$g$ compared to the expected order $e^2 g^2 T^2$.\cite{foot} 
This fact
warrants a re-examination of the diagrams which had been found to 
vanish in the HTL approximation since they may well contribute
at the suppressed order $e^2 g^3 T^2$. Considering 
Fig.~\ref{figsoft}.d, a blown-up view of the effective
four gauge boson vertex reveals that the diagram is in fact
equivalent to the two-loop diagrams of Fig.~\ref{fighard}
with a hard fermion loop and a soft gluon insertion: 
Fig.~\ref{fighard} corresponds to the lowest order
bremsstrahlung diagrams studied in \cite{weldon,cleymans}. The same
reasoning can be made for Fig.~\ref{figsoft}.c and leads to
the diagram with a self energy insertion of Fig.~\ref{fighard}.b
where the gluon is now hard and the fermion of momentum $R+L$
soft. In the following we calculate only the diagrams of
Fig.~\ref{figsoft}.d, beyond the hard loop approximation,
because it is enhanced, compared to Fig.~\ref{figsoft}.c, 
by a factor $1/g$, due to the Bose factor 
associated to the soft gluon propagator.

We work in the Retarded/Advanced formalism of finite temperature
field theory \cite{pat,ra,van,van2}, which has the advantages of real time
methods \cite{fett,niem,kobe} but still maintains close ties with the imaginary time
techniques \cite{joe,land}. 
For the contribution of the graph with the self--energy
correction (we must keep in mind that there is another
graph with a self-energy correction on the other fermionic propagator),
 we find the following result for the retarded self-energy
up to colour and group factors, which will be re--established
in the final results \cite{ra,pat}:
\begin{eqnarray} 
 -{i\Pi^\mu\,_\mu(Q)}_{|_{R}}  &=& - e^2 
\int {d^nP\over(2\pi)^n} \nonumber\\
  & &\times\left\{\left[\frac{1}{2}-n_{_{F}}(p_0)\right]
  \left[\Delta_{_{R}}(P)-\Delta_{_{A}}(P)\right]\left[\Delta_{_{R}}(R)\right]^2
  \hbox{\rm Tr}{\imb\Sigma}_{_{R}} \right.\nonumber\\
  & &+\left. \left[\frac{1}{2}-n_{_{F}}(r_0)\right]\left[
  \Delta_{_{R}}^2(R)\hbox{\rm Tr}{\imb\Sigma}_{_{R}}-
\Delta_{_{A}}^2(R)\hbox{\rm Tr}{\imb\Sigma}_{_{A}}\right]
  \Delta_{_{A}}(P)\right\}
\end{eqnarray}
where the retarded and advanved propagators are defined by
\begin{equation}
 \Delta_{_{R,A}} (P) \equiv {i\over P^2-M^2\pm i\epsilon p_0}.
\end{equation}
(here we use $M=0$) and the notation $\hbox{\rm Tr}{\imb\Sigma}_\alpha$
with $\alpha=R,\,A$ stands for:
\begin{equation}
  \hbox{\rm Tr}{\imb\Sigma}_\alpha\equiv\hbox{\rm Tr}\left(\gamma_\mu \slR
  \left[-i\Sigma_\alpha(R)\right] \slR \gamma^\mu \slP\right).
\end{equation}
The one-loop fermion self-energy $\Sigma_\alpha(R)$ is calculated with
the following decomposition of the soft gluon propagator into 
its transverse, longitudinal and gauge components \cite{rp,brat}:
%
%
\begin{equation}
  \label{prop}
   iD^{\mu\nu}(L)\equiv\frac{P_{_{T}}^{\mu\nu}(L)}{L^2-\Pi_{_{T}}} +
  \frac{P_{_{L}}^{\mu\nu}(L)}{L^2-\Pi_{_{L}}} +\hbox{\rm gauge}\ 
\hbox{\rm terms},
\end{equation}
We introduce the spectral functions 
\begin{equation}
\rho_{_{T,L}}(l, l_0) =  
      \left.{i \over L^2 - \Pi_{_{T,L}} }\right|_{_{R}}
   -   \left.{i \over L^2 - \Pi_{_{T,L}} }\right|_{_{A}}.
\end{equation}
We find that the imaginary part of the photon self-energy,
defined by \\
$ 
2 i\ \hbox{\rm Im} \Pi^\mu\,_\mu = {\Pi^\mu\,_\mu}_{|_R}
 - {\Pi^\mu\,_\mu}_{|_A},
$
can be written at $Q^2=0$ as
\begin{eqnarray}
  \label{self}
 & & \hbox{\rm Im}\,\Pi^\mu\,_\mu(Q)=-2e^2
g^2\int\frac{d^n P}{(2\pi)^{n-1}}
  \epsilon(p_0)\delta(P^2)\left[n_{_{F}}(r_0)-n_{_{F}}(p_0)\right]\nonumber\\
  & &\qquad\times \int\frac{d^n L}{(2\pi)^{n-1}}
  \epsilon(r_0+l_0)\delta\left[(R+L)^2\right]
  \left[n_{_{F}}(r_0+l_0)+n_{_{B}}(l_0)\right]\rho_{_{T,L}}(l, l_0)\nonumber\\
  & &\qquad\times
\left[\frac{4R_\mu Q_\nu P^{\mu\nu}_{_{T,L}}(L)}{R^2}
  -P_{_{T,L}\ \mu}^\mu(L)\left(1+\frac{2Q\cdot L}{R^2}
\right)\right],
\end{eqnarray}
with appropriate spectral function $\rho_{_{T,L}}$ and projection
matrix $P_{_{T,L}}$. 
This expression corresponds to cut ``$b$'' of Fig.~\ref{fighard}.b.  
Cuts ``$a$'' and ``$c$''
vanish in dimensional regularization or, if one uses a
regulating mass for the fermion, by kinematical arguments.
At that point, we drop the gauge dependent part of
the gluon propagator since it is possible to verify that it does not
contribute at the order we are interested in, thus leaving a gauge
independent result.
\par
\begin{figure}[ht]
\begin{center}
\leavevmode
\epsfxsize=5 in
\epsfbox{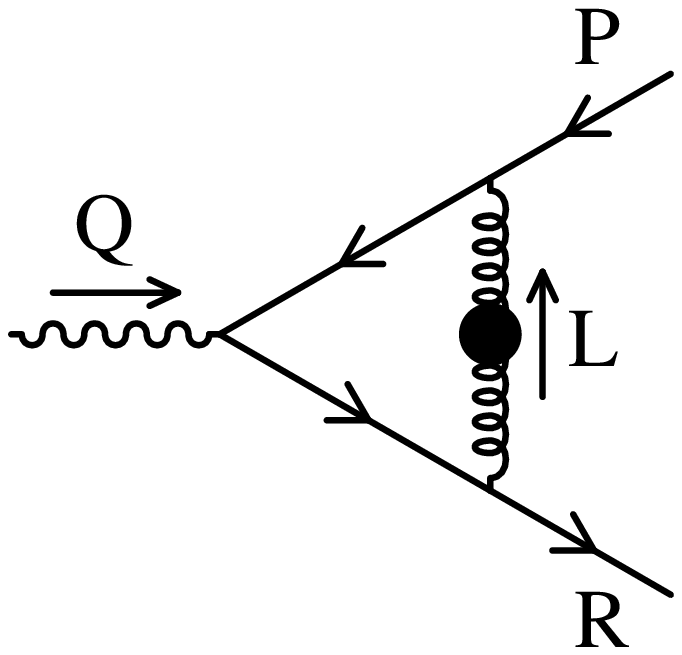}
\end{center}
\caption{The three--point vertex function.}
\label{figvertex}\end{figure}

The contribution of the vertex diagram to the retarded  
self energy can be written as \cite{ra,pat}:
\begin{eqnarray}
& &-i{\Pi^\mu\,_\mu(Q)}_{|_{R}} 
 =  - e^2 
\int {d^nP\over(2\pi)^n} \ \hbox{\rm TR}_{_{T,L}} \nonumber\\
 & & \times\left\{\left[\frac{1}{2}-n_{_{F}}(p_0)\right]
  \left[ \left( V_{_{RRA}}(P,Q,-R)\Delta_{_{R}}(P) - 
          V_{_{ARA}}(P,Q,-R)\Delta_{_{A}}(P) \right) \Delta_{_{R}}(R) 
  \right]
          \right.\nonumber\\
 & & \left. \ \ \ \left[\frac{1}{2}-n_{_{F}}(r_0)\right]
  \left[ \left( V_{_{ARA}}(P,Q,-R)\Delta_{_{R}}(R) - 
          V_{_{ARR}}(P,Q,-R)\Delta_{_{A}}(R) \right) \Delta_{_{A}}(P) 
  \right]
  \right\} 
\label{imvertex}      
\end{eqnarray}
where all the Dirac algebra factors are included in 
\begin{equation}
  \hbox{\rm TR}_{_{T,L}}\equiv\left[P_{_{T,L}}^{\rho\sigma}(L)\right]
\hbox{\rm Tr}\left(\gamma_\mu\slR\gamma_\rho(\slR+\slL)\gamma^\mu(\slP+\slL)
\gamma_\sigma\slP\right)
\end{equation}
and the functions $V_{\alpha \beta \gamma}(P,Q,-R)$ contain the scalar part
of the diagram of Fig.~\ref{figvertex}. They are defined by \cite{ra,pat}:
\begin{eqnarray}
& & V_{\alpha \beta \gamma}(P,Q,-R) \equiv - g^2  
\int {d^nL\over(2\pi)^n} \nonumber \\
& & \left\{ ( \frac{1}{2}+n_{_{B}}(l_0) )
    \left[ \left.{i \over L^2-\Pi}\right|_{_{R}}
          -\left.{i \over L^2-\Pi}\right|_{_{A}} \right] 
     \Delta_\alpha(P+L) \Delta_{\bar\delta}(R+L) \right. \nonumber \\
& & + ( \frac{1}{2}-n_{_{F}}(r_0+l_0) )
   \left[ \Delta_{_R}(R+L) - \Delta_{_A}(R+L) \right]
   \Delta_{\bar\beta}(P+L) \left.\frac{i}{L^2-\Pi}\right|_{\delta} 
   \nonumber  \\
& & \left. + ( \frac{1}{2}-n_{_{F}}(p_0+l_0) )
   \left[ \Delta_{_R}(P+L) - \Delta_{_A}(P+L) \right]
   \Delta_{\beta}(R+L) \left.\frac{i}{L^2-\Pi}\right|_{\bar \alpha} 
   \right\}. 
\label{vert}    
\end{eqnarray}
Plugging this expression in Eq.~(\ref{imvertex}) one obtains the imaginary
part of the self energy:
\begin{eqnarray}
  \label{vertex}
\hbox{\rm Im}\,\Pi^\mu\,_\mu(Q) &=& -\frac{1}{2}e^2g^2
\int\frac{d^n P}{(2\pi)^{n-1}}
  \left[n_{_{F}}(r_0)-n_{_{F}}(p_0)\right] \int\frac{d^n L}{(2\pi)^{n-1}}
  \rho_{_{T,L}}(l, l_0)  \ \hbox{\rm TR}_{_{T,L}}  \nonumber\\
& &\times\left\{\epsilon(p_0)\delta(P^2)
  \epsilon(r_0+l_0)\delta\left[(R+L)^2\right]
  {n_{_{F}}(r_0+l_0)+n_{_{B}}(l_0) \over R^2 (P+L)^2} \right.\nonumber\\
& & \ \ \left.+\epsilon(r_0)\delta(R^2)
  \epsilon(p_0+l_0)\delta\left[(P+L)^2\right]
  {n_{_{F}}(p_0+l_0)+n_{_{B}}(l_0) \over P^2 (R+L)^2} \right\}
\end{eqnarray}
Two classes of terms appear which can be interpreted in terms of cut
diagrams: the first term in the curly brackets above coresponds to cut 
``$b$" in Fig.~\ref{fighard}.a and it is to be combined with Eq.~(\ref{self})
while the second term  is associated with the other
self-energy correction mentioned above. Both classes of terms give
an identical contribution to the photon production rate. 
Keeping only the dominant terms for $P,\,R$ hard and $Q,\,L$ soft, 
we find the total contribution to the imaginary part of the self--energy 
to be
\begin{eqnarray}
 & & \hbox{\rm Im}\,\Pi^\mu\,_\mu(Q)=-2e^2g^2
\int\frac{d^n P}{(2\pi)^{n-1}}
  \int\frac{d^n L}{(2\pi)^{n-1}}
  {q} n_{_{F}}'(p_0)\; n_{_{B}}(l_0)\rho_{_{T,L}}(l,l_0) \nonumber\\
  & &\qquad\times 
  \epsilon(p_0)\epsilon(r_0+l_0)\delta(P^2)
\delta\left[(R+L)^2\right]
  4R_\rho P_\sigma P_{_{T,L}}^{\rho\sigma}
  \frac{L^2}{R^2(P+L)^2}.
\label{sum}
\end{eqnarray}
The presence of the $L^2$ factor in the numerator clearly indicates that our
calculation is carried out beyond the HTL approximation where such terms
are neglected. Estimating naively the order of magnitude of our result
(using $L^2 \sim g^2 T^2$ and $R^2 \sim (P+L)^2 \sim g T^2$) we find it
to be  ${\cal O }(e^2 g^3 T^2) $, i.e. of the same order as the supposedly
dominant soft fermion loop contribution of Eq.~(\ref{soft}). However the
denominator $R^2(P+L)^2$ is responsible for collinear divergences which 
drastically modify this naive estimate. Using the $\delta$ function constraints
one easily rewrites 
\begin{eqnarray}
{-4 \over R^2 (P+L)^2} &=& {1 \over P\cdot Q} {1 \over P\cdot Q + Q\cdot 
L} =  
              {1 \over Q\cdot L} \left({1 \over P\cdot Q} 
                            - {1 \over P\cdot Q + Q\cdot L} \right) \nonumber \\
                      &\approx&  {2 \over Q\cdot L} {1 \over P\cdot Q}.
\label{diverg}
\end{eqnarray}
The first equality shows the presence of two very close collinear singularities
($P\cdot Q$=0) since the two poles differ only by the soft $Q\cdot L$ term. The 
last equality holds 
true to leading order
only after the integration over the whole phase
space in Eq.~(\ref{sum}) is performed. Introducing the angular variable 
$u=1-\cos\theta$ between the light-like momenta $P$ and $Q$ the above expression 
becomes, near $u=0$, 
\begin{eqnarray}
{1 \over R^2 (P+L)^2} \sim  {p \over q L^2}\  {1 \over p q u} 
\label{collin}
\end{eqnarray}
This form shows the presence of a logarithmic collinear divergence  
and the order of the residue at the pole in $u$ is 
$1 / g^4 T^4$ instead of the naively expected $1 / g^2 T^4$.
Concentrating then on this collinear limit, and leaving details
of the calculations to a future paper, we find the dominant 
divergent term to be
\begin{eqnarray}
  \label{massless}
  & &\hbox{\rm Im}\,\Pi^\mu\,_\mu(Q)\approx(-1)_{_{L}} 
4e^2  g^2 
N_{_{C}}C_{f}
\frac{1}{(2\pi)^4}
  \frac{1}{q}\int_{p^*}^\infty dp\,p^2 n_{_{F}}(p)(1-n_{_{F}}(p))\nonumber\\
& &\qquad\times\int_0^{l^*}\frac{dl}{l}
  \int_{-l}^{l}\frac{dl_0}{l_0}L^2\rho_{_{L,T}}(l_0,l)
  \quad \int_0^1 \frac{du}{u},
\end{eqnarray}
where $p^*$ and $l^*$ are some intermediate momenta
between the hard and soft scale. We have re-introduced at that point
the color factor $N_{_{C}}$ and group factor $C_{f}$ in the result.
 The symbol $(-1)_{_L}$ is
$+1$ for the transverse gluon mode and $-1$ for the longitudinal one.
We find in the $l$ integration that the 
region between $l^*$ and $\infty$ gives a negligible
contribution, so we can take $l^*\to\infty$.
Similarly, we can take $p^*\rightarrow 0$.
We notice a nice factorization of Eq.~(\ref{massless})
 into a ``hard thermal loop''
integral over $p$, a soft gluonic integral over $l$ and $l_0$,
and an integral over the angular variable $u$ leading to the 
logarithmic collinear divergence. 
It is possible to show by kinematical considerations
that only the Landau damping part of the spectral function 
$\rho_{_{L,T}}$ contributes to the divergent piece; this is the
reason why the integration domain has been limited to $L^2<0$.
The occurence of a collinear divergence, as in Eq.~(\ref{massless}),
was noticed in the dispersion relations
for scalar $QED$ near the light cone \cite{kraemmer}.

\par

To proceed, sum rules may then be used to reduce the
integrations of Eq.~(\ref{massless}) down to a single one;
for example, for the
transverse contribution we can use
\begin{eqnarray}
 & &\frac{1}{\pi}\int_{-\infty}^{+\infty}dz\,z\,\rho_{_{T}}(zl,l)=
  \frac{2}{l^2},\nonumber\\
  & &\frac{1}{\pi}\int_{-\infty}^{+\infty}\frac{dz}{z}
  \rho_{_{T}}(zl,l)=\frac{2}{l^2},\nonumber\\
   & &\frac{1}{\pi}\int_{-1}^{+1}dz\,z\,\rho_{_{T}}(zl,l)=
  \frac{2}{l^2}[1-Z_{_{T}}(l)],\nonumber\\
  & &\frac{1}{\pi}\int_{-1}^{+1}\frac{dz}{z}\rho_{_{T}}(zl,l)=2\left[
  \frac{1}{l^2+m_{\hbox{\cmr mag}}^2}-\frac{Z_{_{T}}(l)}{\omega_{_{T}}^2(l)}\right],
\end{eqnarray}
where
\begin{eqnarray}
  & &Z_{_{T}}\equiv\frac{2\omega_{_{T}}^2(\omega_{_{T}}^2-l^2)}
  {3m_g^2\omega_{_{T}}^2-(\omega_{_{T}}^2-l^2)^2},
\end{eqnarray}
where $\omega_{_{T}}(l)$ is the energy of the solution to the
dispersion relation and $m_g$ is the gluon Debye mass.
Note that we have
introduced a phenomenological ``magnetic mass''
$m_{\hbox{\cmr mag}}\sim g^2T$ by hand to regulate potential infrared
divergence for the transverse contribution. 
We find the divergent piece of the transverse term
can be written as
\begin{eqnarray}
 \hbox{\rm Im}\,\Pi^\mu\,_\mu(Q)&\approx&
  \frac{e^2 g^2 N_{_{C}} C_f}{12\pi}
  \frac{T^3}{q}\frac{1}{\varepsilon}
  \int_0^\infty \frac{dl}{l}\left\{
\frac{m_{\hbox{\cmr mag}}^2}{l^2+m_{\hbox{\cmr mag}}^2}-
    Z_{_{T}}\frac{\omega_{_{T}}^2-l^2}{\omega_{_{T}}^2}
    \right\}
\nonumber\\
  &\sim& O(e^2gT^2)\frac{1}{\varepsilon}.
\end{eqnarray}
Similar sum rules can be used to evaluate the longitudinal contribution 
which is found to be
\begin{eqnarray}
 \hbox{\rm Im}\,\Pi^\mu\,_\mu(Q)&\approx&
  -\frac{e^2  g^2N_{_{C}} C_f}{12\pi}
  \frac{T^3}{q}\frac{1}{\varepsilon}
  \int_0^\infty \frac{dl}{l}\left\{
\frac{3m_g^2}{l^2+3m_g^2}-
    Z_{_{L}}\frac{\omega_{_{L}}^2-l^2}{\omega_{_{L}}^2}
    \right\}
\nonumber\\
  &\sim& O(e^2gT^2)\frac{1}{\varepsilon}.
\end{eqnarray}
We note that the transverse contribution requires
the ``magnetic mass'' introduced earlier
in the sum rules in order to be infrared safe but both contributions 
display a divergence which is seen to be 
enhanced by a factor of $1/g^2$ relative to
the soft fermion loop contribution of Eq.~(\ref{soft}).
\par
In our discussion of the collinear divergences care has been taken to keep
the exact kinematics in the evaluation of the denominators: thus, in 
Eq.~(\ref{diverg}), the ``soft" term $Q\cdot L$ was not neglected compared
to the ``hard" term $P\cdot Q$. But there are other soft corrections to hard
propagators which should also be included, namely those associated to
thermal mass effects. Taking these into account amounts to apply
a further resummation of hard internal lines
\cite{kraemmer,flech}. This resummation is in addition to the HTL 
resummation of Braaten and Pisarski for soft lines, and is important
for processes that are sensitive to the behaviour near the light cone.
It also has the virtue of being a gauge invariant resummation
summarized by a compact effective action.
In the present case this involves using the dressed fermion propagator
given in the limit $p_0,\,p \gg gT$ by
\begin{equation}
  \frac{P_0\gamma^0-\omega_+(p)\hat{P}\cdot\vec{\gamma}}
  {P^2-M_\infty^2+O(M_\infty^4/p^2)}\approx
  \frac{\slP}{P^2-M_\infty^2},
\end{equation}
where $M_\infty$ is the fermionic thermal mass in the hard regime and 
$\omega_+(p)\approx\surd(p^2+M_\infty^2)$ is the energy of the 
``particle
mode'' of the fermionic dispersion relation.
Furthermore, this asymptotic mass is insensitive to a soft
modification of the hard propagator. 
Carrying through the calculations with such
a propagator, one finds the analagous relation to the final result
of Eq.~(\ref{massless}) as
\begin{eqnarray}
  \label{massive}
  & &\hbox{\rm Im}\,\Pi^\mu\,_\mu(Q)\approx(-1)_{_{L}} 
  4e^2  g^2 N_{_{C}} C_f 
\frac{1}{(2\pi)^4}
  \frac{1}{q}\int_{p^*}^\infty dp\,p^2 n_{_{F}}(p)(1-n_{_{F}}(p))
\nonumber\\
& &\qquad\times\int_0^{l^*}\frac{dl}{l}
  \int_{-l}^{l}\frac{dl_0}{l_0}L^4\rho_{_{L,T}}(l_0,l)
 \quad \int_0^1 \frac{du}{\sqrt{1-u}}
  \frac{1}{L^2u-4M_\infty^2}.
\end{eqnarray}
\par
We note that taking the $M_\infty\to 0$
limit of Eq.~(\ref{massive}) at the collinear point
results in Eq.~(\ref{massless}), reproducing the collinear
divergence as well as the dependence on the magnetic mass.
The angular integration in Eq.~(\ref{massive}) is easily performed and one
arrives at
\begin{eqnarray}
& & \hbox{\rm Im}\,\Pi^\mu\,_\mu(Q)\approx(-1)_{_{L}}
  \frac{e^2 g^2 N_{_{C}} C_f}{12\pi}
  \frac{T^3}{q}\frac{2}{\pi}\int\limits_{0}^{1}\;
\frac{dx}{x}\;\tilde{I}_{_{T,L}}(x) \nonumber\\
& & \qquad\times\int\limits_{0}^{+\infty}\;dw\;
\frac{\sqrt{w/(w+4)}
\hbox{\rm tanh}{}^{-1}\sqrt{w/(w+4)}}{(w+\tilde{R}_{_{T,L}}(x))^2+
(\tilde{I}_{_{T,L}}(x))^2}
\end{eqnarray}
with $w = -L^2/M_\infty^2$, $\tilde{R}_{_{T,L}}(x)\equiv\hbox{\rm Re}\,\Pi_{_{T,L}}(x)/M_\infty^2$ and 
$\tilde{I}_{_{T,L}}(x)\equiv\hbox{\rm Im}\,\Pi_{_{T,L}}(x)/M_\infty^2$.
Obviously the remaining integrals are finite and dimensionless.
We note, though, that as in the massless case of
Eq.~(\ref{massless}) that sum rules can be used
to reduce Eq.~(\ref{massive}) down to a one--dimensional
integral.
Details will be given in a future
work; we simply describe here some general
features about the
 result which are evident from this form
and from Eq.~(\ref{massive}). The first is
that the order of the contributions is not
changed by inclusion of the asymptotic mass, and as such, it is
also enhanced by a factor of $1/g^2$ relative to the
soft contribution of Eq.~(\ref{soft}).
The second one is
that the presence of $M_\infty$ regulates the
collinear divergence associated with the lower limit on the 
$u$ integration, since $L^2<0$. The final one is
 that the former
sensitivity on the magnetic mass scale disappears 
when one considers specifically 
$M_\infty \sim {\cal O}(gT)$, but, as noted after
Eq.~(\ref{massive}) in taking $M_\infty\to 0$,
would reappear at scales below this.
This result is thus of the same order as that of 
Ref.~\cite{cleymans},
neglecting the Landau-Pomeranchuck effect, but we differ
from it since we find that both the transverse and the longitudinal
modes contribute to the same order while in Ref.~\cite{cleymans}
the interaction in the medium is assumed to be static.

\section{Conclusions}

Although this resummation of the hard fermion line by
inclusion of the
asymptotic mass regulates the collinear divergence,
we should note
that it is not rigorously known if such a mass is the only term
present at this
order in general. In particular, terms in an effective
propagator
which might arise from higher loop diagrams than the
hard thermal one loop
terms giving rise to this asymptotic mass may contribute.
Thus, we cannot
say with absolute certainty that the terms discussed here are
the only
ones which contribute at this order, although arguments exist
that this may in fact be the case.
Indeed, in this regard cancellations
may occur: it is known in some examples that, for example, a constant
damping term inserted in an effective propagator will cancel
against the corresponding vertex corrections \cite{smilga,flech,meg}.
What does seem clear, though,
is that
there is an enhancement mechanism present in processes near
the light
cone which falls outside of the usual Braaten--Pisarski
resummation of
soft internal lines. It is also possible through a similar
mechanism that other processes sensitive to the behaviour
of the theory near the light cone, such as the photon
production rate for slightly virtual photons \cite{yuan},
may also get contributions from similar terms with hard
internal momenta. Work along these lines, as well as
details of the calculations reported here, will be presented
elsewhere.

\section{Acknowlegements}

We thank R.~Baier and A.~Rebhan for valuable
discussions. The work of PA
is supported in part by the EEC program
``Human Capital and Mobility", Network 
``Physics at High Energy Colliders", 
contract CHRX-CT93-0357 (DG 12 COMA).
The work of RK and EP is supported by
the Natural Sciences and Engineering Research Council
of Canada.

\par\section*{Figure Captions}
\begin{itemize}
\item[Fig.~1] Contributions to the soft photon production
rate with soft internal lines.
\item[Fig.~2] Contributions to the soft photon production
rate with hard internal fermion lines -- (a): vertex insertion;
(b): self--energy insertion.
\item[Fig.~3] The three--point vertex function.
\end{itemize}
\clearpage
\clearpage
\clearpage

\begin{thebibliography}{99}
\bibitem{foot} It is worth recalling, in this respect, that the trace of the 
polarization tensor has no hard thermal loop contribution while
the transverse and the longitudinal polarization tensors each have
a HTL contribution.
\end{thebibliography}
\end{document}